\documentclass[12pt,twoside]{article}
\usepackage[mathscr]{eucal}
\usepackage{amsmath,amsfonts,amssymb,amsthm}

\usepackage[matrix,arrow]{xy}
\usepackage{times}
\usepackage{graphicx}
\usepackage{epstopdf}

\advance\textheight\topskip
\renewcommand{\tilde}{\widetilde}
\renewcommand{\hat}{\widehat}
\newtheorem{prop}{Proposition}[section]

\newtheorem{theorem}[prop]{Theorem}

\renewcommand{\d}{\partial}
\def\ndelta{\delta\hspace{-0.50em}\slash\hspace{-0.05em} }

\newcommand{\half}{\frac{1}{2}}

\newcommand{\RR}{\mathbb{R}}

\def\cG{\mathcal{G}}

\def\cL{\mathcal{L}}
\def\cM{\mathcal{M}}

\def\cQ{\mathcal{Q}}

\numberwithin{equation}{section} \makeatletter
\@addtoreset{equation}{section}

\begin{document}

\def\mytitle{Enhanced asymptotic symmetry algebra of $AdS_3$.}

\pagestyle{myheadings} \markboth{\textsc{\small Troessaert}}{%
  \textsc{\small Enhanced symmetry algebra of $AdS_3$}} \addtolength{\headsep}{4pt}

\begin{centering}

  \vspace{1cm}

  \textbf{\Large{\mytitle}}



  \vspace{1.5cm}

  {\large C\'edric Troessaert}

\vspace{.5cm}

\begin{minipage}{.9\textwidth}\small \it \begin{center}
   Centro de Estudios Cient\'ificos (CECs)\\ Arturo Prat 514,
   Valdivia, Chile \\ troessaert@cecs.cl \end{center}
\end{minipage}

\end{centering}

\vspace{1cm}

\begin{center}
  \begin{minipage}{.9\textwidth}
    \textsc{Abstract}. A generalization of the Brown-Henneaux boundary
    conditions is introduced for pure gravity with negative
    cosmological constant in 3 dimensions. This leads to new degrees of
    freedom and to an
    enhancement of the symmetry algebra. Up to the zero modes, it consists of
    two copies of the semi-direct product of a Virasoro algebra with a
    $U(1)$ current algebra. The 
    associated surface charge algebra now contains three non-zero central charges:
    the two usual Brown-Henneaux central charges and one new quantity.
  \end{minipage}
\end{center}

\thispagestyle{empty}
\newpage


\section{Introduction}
\label{sec:introduction}

Einstein's gravity in 2+1 dimensions is an interesting toy
model to understand some features of higher dimensional gravity as, in
three dimensions,
this theory doesn't have local degrees of freedom but still has
dynamical global objects \cite{Deser:1983nh}. 

For instance, in the presence of a negative cosmological
constant, there exists the famous BTZ black-hole solution
 \cite{Banados:1992fk, Banados:1993uq}. Those black-holes possess
the same characteristics as their higher dimensional 
cousins like temperature or entropy. One hopes that understanding
the BTZ black-holes thermodynamical properties in this simpler setup
would help us in the more physically relevant cases.

Another surprise came even earlier with the study by Brown-Henneaux of
the symmetry algebra of asymptotically $AdS_3$ space-times \cite{Brown:1986zr}. They showed that this algebra
is not the expected $so(2,2)$ symmetry algebra of the background
$AdS_3$ but is enhanced to the full conformal algebra in two
dimensions. Furthermore the algebra
acquires classical central charges with the famous value of
$c^\pm=\frac{3l}{2G}$. Having the full 2D conformal group as a symmetry of the theory allows
for the use of the powerful techniques of 2D CFT's. One of the
main results is then the computation by Strominger where 
he was able to reproduce the Bekenstein-Hawking entropy of the BTZ
black-holes using the Cardy formula and the explicit value of the
Brown-Henneaux central charges \cite{Strominger:1998fk}. 

In the last fifteen years, a lot has been done to further improve our
understanding. For instance, using the Chern-Simons description of gravity in 3D, it was shown
that gravity in this case is
equivalent to a Liouville theory on the boundary
\cite{Coussaert:1995zp, Henneaux:1999ib, Rooman:2000zi, Barnich:2013yka}. However,
Liouville theory does not contain enough degrees of freedom to fully
account for the entropy of the black-holes \cite{Banados:1998ys, Carlip:2005zn}. More recently, a direct
computation of the partition function of the theory was done but, in most cases, the
results are not sensible \cite{Maloney:2007ud, Castro:2011zq}. Those
are some of the more recent results but, in general, we still don't
have a description of the fundamental degrees of freedom of the
theory.

All the results described above are strongly dependent on the
Brown-Henneaux boundary conditions and the resulting asymptotic
symmetry algebra. Attempts have been made to try to relax them but
with no impact on the number of global degrees of
freedom and no change on the asymptotic symmetry algebra
\cite{Porfyriadis:2010vg}. A few days ago, the authors of
\cite{Compere:2013bya} proposed a new set of chiral boundary
conditions for asymptotically $AdS_3$ space-times. Those new conditions
are associated to a different problem as they only contain part of the
solutions to Einstein's equations satisfying to the Brown-Henneaux
boundary conditions.

In this paper, we want to present a set of boundary conditions that
generalizes the one of Brown-Henneaux. Those boundary conditions
describe a theory with more degrees of freedom. Moreover, there is a
second enhancement of the asymptotic symmetry algebra. Up to the zero modes, the new algebra is generated by two Virasoro algebras and two $U(1)$ current algebras. At the level of the charges, the
algebra acquires a central extension characterized by three non-zero numbers: the two usual Brown-Henneaux
central charges plus one new quantity.

\section{New asymptotic conditions}
\label{sec:new-asympt-cond}

The action for gravity in 3 dimension with cosmological constant is
the Einstein-Hilbert action:
\begin{equation}
S[g] = \frac{1}{16 \pi G}\int_\cM d^3x\,  \left(R - 2
  \Lambda\right), \quad \Lambda= - \frac{1}{l^2}, \quad \cM = \RR^3.
\end{equation}
This action is not well defined without additional boundary terms
and fall-off conditions for the fields. The usual setup is given by the
Brown-Henneaux boundary conditions \cite{Brown:1986zr}:
\begin{eqnarray}
\label{eq:BHasymp1}
g_{AB} & = & r^2 \bar \gamma_{AB} + O(1),\\
g_{rA} & = & O(r^{-3}),\\
\label{eq:BHasymp3}
g_{rr} & = & \frac{l^2}{r^2} + O(r^{-4}),
\end{eqnarray}
where $\bar \gamma_{AB}$ is a fixed metric on the cylinder at spatial
infinity $(r \rightarrow \infty)$ and $x^A=(\tau, \phi)$. Choosing
$\bar \gamma_{AB}$ as the flat metric corresponds to asymptotically $AdS_3$
space-times.  With the metric $\bar \gamma_{AB}$ fixed, the action can be 
supplemented with the Gibbons-Hawking boundary term to make it well
defined \cite{Gibbons:1976ue,Brown:1993fk}:
\begin{equation}
\label{eq:fullaction}
S[g] = \frac{1}{16 \pi G}\int_\cM d^3x\, \sqrt{-g} \left(R - 2 \Lambda\right) + \frac{1}{16 \pi G}\oint_{\d
  \cM} d^2x \,\sqrt{-h}\left(-2K +{}_0 K\right),
\end{equation}
where $h_{AB}=g_{AB}$ is the induced metric and
$K=h^{AB}K_{AB}$ is the trace of the 
extrinsic curvature of the boundary. We only considered the time-like part of $\d \cM$ wich
is the cylinder at spatial infinity. The
quantity ${}_0K = \frac{-2}{l}$ acts as a counter-term to make the
boundary stress energy tensor finite \cite{Henningson:2000vn,Balasubramanian:1999uq,deHaro:2000wj}.

We will argue that a more general possibility is to use the same asymptotic behavior
\eqref{eq:BHasymp1}-\eqref{eq:BHasymp3} but fixing only the conformal structure of the
induced metric on the boundary \cite{Papadimitriou:2005dq,Compere:2008us}:
\begin{equation}
g_{AB} = r^2 \gamma_{AB} + O(1), \qquad \gamma_{AB} = e^{2\varphi} \bar \gamma_{AB}
\end{equation}
where $\varphi$ will be a dynamical field. Varying the action
\eqref{eq:fullaction} now leads to
\begin{multline}
\delta S[g] = \frac{1}{16 \pi G}\int_\cM d^3x\, \sqrt{-g} \delta
g_{\mu\nu} \left(-G^{\mu\nu} -\Lambda g^{\mu\nu}\right)\\ + \frac{1}{16 \pi G}\oint_{\d
  \cM} d^2x \, r^2\sqrt{-\gamma} \, 2 \delta \varphi \left( -K +{}_0K \right).
\end{multline}
In order to have a well
defined action, we need to impose $K+\frac{2}{l}=o(r^{-2})$. This corresponds to a simple
change from Dirichlet to Mixed boundary conditions. On
space-like boundaries, it is equivalent to a canonical transformation
but as we will see, on time-like boundaries, it changes the number of
global degrees of freedom drastically. 

\vspace{3mm}

The asymptotic conditions discussed above can be
summarized by:
\begin{eqnarray}
\label{eq:condasymp1}
g_{rr} &= & \frac{l^2}{r^2} + C_{rr} r^{-4}+o(r^{-4}),\\
g_{rA} & = & O(r^{-3}),\\
\label{eq:condasymp3}
g_{AB} & = & r^2\gamma_{AB}+ C_{AB}+o(1),\\
\label{eq:condasymp4}
K - {}_0K &=& o(r^{-2}),
\end{eqnarray}
where $\gamma_{AB} = e^{2\varphi} \bar\gamma_{AB}$ and $\bar\gamma$ is a
fixed metric on the cylinder. The last condition is a
constraint on the functions $\varphi(x^A), C_{rr}(x^A), C_{AB}(x^C)$:
\begin{equation}
\label{eq:suppcond}
\rho \equiv \gamma^{AB}C_{AB} + \frac{1}{l^2} C_{rr} = 0.
\end{equation}

\vspace{3mm}

To describe asymptotically $AdS_3$ space-times, the metric $\bar
\gamma$ has to be fixed to the flat metric:
\begin{equation}
\bar \gamma_{AB} dx^Adx^B = -d\tau^2 + d\phi^2 = -dx^+ dx^-
\end{equation}
where $x^\pm=\tau\pm\phi$ are light-cone coordinates on the
cylinder. The BTZ black-hole \cite{Banados:1992fk, Banados:1993uq}
satisfies those asymptotic conditions with $\varphi=0$ and
\begin{equation}
C_{\tau\tau}=8Gl^2 M,\quad  C_{\phi\phi} = 0, \quad C_{\tau\phi} = 4Gl
J \quad \text{and} \quad C_{rr} = 8Gl^4 M,
\end{equation}
or, in light-cone coordinates:
\begin{equation}
C_{+-} = 2G l^2 M, \quad C_{\pm\pm} = 2Gl^2 \left(M \pm \frac{J}{l} \right).
\end{equation}
As we will see in section \ref{sec:asymptotic-solutions}, any solution
of Einstein's equation satisfying the Brown-Henneaux boundary
conditions \eqref{eq:BHasymp1}-\eqref{eq:BHasymp3} also satisfy the
supplementary condition $\rho = 0$. In that sense, those new boundary  conditions
\eqref{eq:condasymp1}-\eqref{eq:condasymp4} are a
generalization of the usual ones. 

\section{Asymptotic symmetries}
\label{sec:asympt-sym}

The infinitesimal diffeomorphisms leaving the conditions \eqref{eq:condasymp1}-\eqref{eq:condasymp4} invariant are
generated by vector fields $\xi^\mu$ satisfying
\begin{gather}
\cL_\xi g_{rr}=\eta_{rr}r^{-4} + o(r^{-4}),\quad \cL_\xi g_{rA}=
O(r^{-3}), \label{eq:asympIa} \\\label{eq:asympIb}\quad
\cL_\xi g_{AB}=2 \omega \, \gamma_{AB} r^2 + \eta_{AB}+o(1),\\
-2\omega\, \gamma^{AB}C_{AB} +\gamma^{AB}\eta_{AB} + \frac{1}{l^2}
\eta_{rr} = 0, \label{eq:asympII}
\end{gather}
where $\cL_\xi$ is the Lie derivative. For convenience, we have
denoted the induced variations on $C_{AB}$, $C_{rr}$ and $\varphi$ by
$\eta_{AB}$, $\eta_{rr}$ and $\omega$ respectively. Equation
\eqref{eq:asympII} is coming from the variation of the supplementary
condition \eqref{eq:suppcond}.
Equations \eqref{eq:asympIa}-\eqref{eq:asympIb} lead to
\begin{eqnarray}
\left\{\begin{array}{l}\xi^r =-\half\psi r + O(r^{-1}), \\ \xi^A  =  Y^A 
   -\frac{l^2}{4r^2}
\gamma^{AB}\d_B
  \psi+O(r^{-4}),
\end{array}\right.\label{eq:FGvect}
\end{eqnarray}
where $Y^A$ is a conformal Killing vector of $\gamma_{AB}$. The last
equation \eqref{eq:asympII} implies that 
$\psi$ is a harmonic function: a solution of $ \Delta \psi=  D_A D^A
\psi=0$ where the derivative $ D_A$ is the covariant derivative
associated with $\gamma_{AB}$. Those conditions on $Y^A$ and $\psi$
depend only on the conformal structure of the cylinder: $\bar\gamma_{AB}$.

We expect the above vectors to form a closed algebra under the Lie
bracket. As in \cite{Barnich:2010fk}, the vectors explicitly depend on the
dynamical part of the metric $g_{\mu\nu}$. In that case, the usual Lie
bracket has to be modified to take into account this dependance. The
relevant bracket is given by: 
\begin{equation}
  \label{eq:modbracket}
 [\xi_1,\xi_2]^\mu_M =[\xi_1,\xi_2]^\mu-\delta^g_{\xi_1}\xi^\mu_2+
\delta^g_{\xi_2}\xi^\mu_1,
\end{equation}
where we denoted by
$\delta^g_{\xi_1}\xi^\mu_2$ the change induced in $\xi^\mu_2(g)$ due
to the variation $\delta^g_{\xi_1}g_{\mu\nu}=\cL_{\xi_1}g_{\mu\nu}$.
Under this bracket, the vectors \eqref{eq:FGvect} satisfy:
\begin{equation}
\begin{array}{rclrcl}
\left[\xi_1,\xi_2\right]^r_M & = & -\half\hat\psi r + O(r^{-1}), &
\hat \psi&=&Y_1^A \d_A \psi_2 - Y_2^A \d_A \psi_1,\\
\left [\xi_1,\xi_2 \right]^A_M& = & \hat Y^A -\frac{l^2}{4r^2}
\gamma^{AB}\d_B\hat\psi+O(r^{-4}), & \hat
Y^A&=&Y^B_1\d_BY_2^A-Y^B_2\d_BY_1^A.
\end{array}
\end{equation}
One can easily prove that
$\hat \psi$ is again a solution of $\Delta \hat \psi=0$. 

The set of transformations for which $Y^A=0=\psi$ is an ideal of the
full algebra. This is the subalgebra of pure gauge transformations; as we
will see in section~\ref{sec:surface-charges}, their associated
charges are zero. The asymptotic symmetry algebra is defined as the
quotient of the full algebra given by \eqref{eq:FGvect} with the ideal
of the pure gauge transformations \cite{Regge:1974kx,Benguria:1977fk}. This quotient is parametrized by
$(Y^A, \psi)$ and the induced Lie bracket is
\begin{equation}
\label{eq:asymptalg}
\left[(Y_1,\psi_1), (Y_2, \psi_2) \right] = (Y^B_1\d_BY_2^A-Y^B_2\d_BY_1^A, Y_1^A \d_A \psi_2 - Y_2^A \d_A \psi_1).
\end{equation}
This algebra is the semi-direct product of the two dimensional conformal
algebra with the harmonic Weyl transformations. It is a subalgebra of
the Penrose-Brown-Henneaux algebra introduced in
\cite{Imbimbo:2000ve} which is the semi-direct product of the two dimensional conformal
algebra with all Weyl transformations.

\vspace{3mm}

In the case of asymptotically $AdS_3$ space-times $\bar \gamma_{AB}
dx^Adx^B = -dx^+dx^-$, the conformal Killing equation for $Y^A$ gives as usual $Y^+(x^+)$ and
$Y^-(x^-)$. The harmonic equation for $\psi$ takes the form
\begin{equation}
\Delta \psi = -4 e^{-2 \phi }\d_+ \d_- \psi.
\end{equation}
Using Fourier expansion, we easily find the general solution:
\begin{equation}
\psi = \sum_{n} \left( \psi^+_n e^{in x^+} + \psi^-_n e^{inx^-}
\right) + V\tau,
\end{equation}
where $\psi^\pm_n$ and $V$ are constants.
Denoting $W^\pm(x^\pm) = \sum_{n} \psi^\pm_n e^{in x^\pm}$, the
algebra \eqref{eq:asymptalg} takes the form:
\begin{eqnarray}
\hat Y^\pm & = & Y_1^\pm \d_\pm Y_2^\pm - Y_2^\pm \d_\pm Y_1^\pm,\\
\hat W^\pm & = & Y_1^\pm (\d_\pm W_2^\pm + \frac{1}{2} V_2) - Y_2^\pm (\d_\pm
W_1^\pm + \frac{1}{2} V_1),\\
\hat V & = & 0.
\end{eqnarray} 

In terms of the basis vectors $l^\pm_n$, $p^\pm_n$ and $q$ defined as
\begin{eqnarray}
\label{eq:asympgenI}
Y^\pm(x^\pm)\d_\pm=\sum_{n\in \mathbf Z} c^n_{\pm} l^\pm_n,&\qquad&
l^\pm_n=e^{inx^\pm}\d_{\pm} , \\
\label{eq:asympgenII}
W^\pm=\sum_{n = 0} \psi^\pm_n p^\pm_n,&\qquad&
p^\pm_n=e^{inx^\pm}, \\
\psi = W^+ + W^- +V q, & \qquad & q=\tau,
\end{eqnarray}
the algebra reads 
\begin{equation}
\boxed{\begin{array}{rclcrcl}
i\left[l^\pm_m,l^\pm_n\right]&=&(m-n)l^\pm_{m+n}, & \qquad & i[l^+_m,l^-_n]&=&0, \\
i\left[l^\pm_m,p^\pm_n\right]&=&-n p^\pm_{m+n}, & \qquad &  i[l^\pm_m,p^\mp_n]&=&0, \\
i\left[p^\pm_m,p^\pm_n\right]&=&0, & \qquad & i[p^+_m,p^-_n]&=&0, \\
i\left[l^\pm_m,q\right]&=&\frac{i}{2} p^\pm_m, & \qquad & i[p^\pm_m,q]&=&0.
\end{array}}\label{eq:ads3alg}
\end{equation}
The two generators $p^+_0$ and $p^-_0$ are identical.
Each chiral copy $(l_m, p_m)$ is the semi-direct product of a Virasoro
algebra with a current algebra. One copy of this semi-direct product
already appeared in the study of asymptotically warped
$AdS_3$ \cite{Henneaux:2011hv,Compere:2009zj,Detournay:2012pc} and in
the study of the new chiral boundary conditions for $AdS_3$ \cite{Compere:2013bya}.

\section{Asymptotic solutions to the EOM}
\label{sec:asymptotic-solutions}

We will solve
Einstein's equations asymptotically for metrics of the form
\eqref{eq:condasymp1}-\eqref{eq:condasymp3} with the last constraint \eqref{eq:suppcond} only
 imposed at the end (see \cite{NavarroSalas:1998ks} for
a similar analysis). To do that, it is useful
to introduce explicitly the first order of $g_{rA}$:
\begin{eqnarray}
\label{eq:Asympsol1}
g_{rr} &= & \frac{l^2}{r^2} + C_{rr} r^{-4}+o(r^{-4}),\\
g_{rA} & = & C_{rA}r^{-3}+ o(r^{-3}),\\
\label{eq:Asympsol3}
g_{AB} & = & r^2\gamma_{AB}+ C_{AB}+o(1).
\end{eqnarray}
For those metrics, the Ricci tensor takes the following form
\begin{eqnarray}
R_{rr} &=& -\frac{2}{l^2} \left( \frac{l^2}{r^2} + C_{rr} r^{-4}\right)
+ o(r^{-4}),\\
R_{rA} & = & \left( -\gamma^{CB}D_B C_{CA} + \gamma^{CB}D_A C_{BC} +
  \frac{1}{2l^2} \d_A C_{rr}\right)r^{-3} \nonumber \\ && \quad - \frac{2}{l^2} C_{rA}r^{-3} +
o(r^{-3}),\\
R_{AB} & = & -\frac{2}{l^2} \left( r^2 \gamma_{AB} +
  C_{AB}\right)\nonumber \\ && \quad +
{}^\gamma R_{AB} + \frac{1}{l^2} \gamma_{AB}\left( \gamma^{CD}C_{CD} +
\frac{1}{l^2}C_{rr}\right) + o(1),
\end{eqnarray}
where ${}^\gamma R_{AB}$ is the Ricci tensor associated to the metric $\gamma_{AB}$.
The EOM $G_{\mu\nu} - \frac{1}{l^2}g_{\mu\nu} = 0$ reduce
asymptotically to two
simple conditions:
\begin{gather}
{}^\gamma R  =  -\frac{2}{l^2} \left( \frac{1}{l^2}C_{rr} + \gamma^{BC}C_{BC}\right) ,\\
D_B (\gamma^{BC} C_{CA} - \frac{1}{2} \delta^B_A  \gamma^{CD} C_{CD})  =  \frac{1}{2}\d_A \left(
  \frac{1}{l^2}C_{rr} + \gamma^{BC}C_{BC} \right).
\end{gather}
They are easily rewritten
in term of $\varphi$ and $\bar \gamma_{AB}$:
\begin{eqnarray}
\label{eq:EOM1}
\bar \Delta \varphi & = & \frac{1}{2}\bar R  +\frac{2 e^{2\varphi}}{l^2} \rho ,\\
\label{eq:EOM2}
\bar D_B \Xi^B_A & =  & \frac{e^{2\varphi}}{2}\d_A \rho, \qquad
\Xi^B_A \equiv \bar\gamma^{BC} C_{CA} - \frac{1}{2} \delta^B_A
\bar\gamma^{CD} C_{CD} 
\end{eqnarray}
where the barred quantities refer to the metric $\bar
\gamma_{AB}$. The quantity $\Xi^B_A $ is a symmetric trace-less
tensor. For our asymptotic conditions, we have to add the constraint
$\rho =0$. This gives us equations of motion for $\varphi$ and
$\Xi^A_B$:  
\begin{equation}
\label{eq:EOMfinal}
\bar \Delta \varphi = \frac{1}{2}\bar R , \qquad \bar D_B \Xi^B_A = 0.
\end{equation}

Using a pure gauge transformation, one can always send a metric
satisfying \eqref{eq:Asympsol1}-\eqref{eq:Asympsol3} to the
Fefferman-Graham gauge-fixed form where $g_{rr}=\frac{l^2}{r^2}$ and $g_{rA} =
0$
\cite{Fefferman:1985uq,Graham:1991kx}. In that case, one can show that Einstein's equations impose 
\begin{equation}
g_{AB} = r^2 \gamma_{AB} + \tilde C_{AB} + r^{-2} S_{AB},
\end{equation}
where $\tilde C_{AB}$ and $S_{AB}$ are given in term of $\varphi$ and
$\Xi^A_B$\cite{Skenderis:2000zr,Graham:1999ly,Rooman:2000qf,Bautier:2000bh,Barnich:2010fk}. In
that sense, $\varphi$ and $\Xi^B_A$ satisfying 
\eqref{eq:EOMfinal} contain all the 
gauge invariant degrees of freedom of the theory. 

\vspace{3mm}

In the usual Brown-Henneaux boundary conditions, one doesn't impose
$\rho=0$ but instead $\gamma_{AB}dx^Adx^B = -d\tau^2 +
d\phi^2$ which implies $\varphi=0$ and $\bar R = 0$. Inserting this in
the full EOM \eqref{eq:EOM1}-\eqref{eq:EOM2}, we obtain:
\begin{equation}
\rho = 0, \qquad \bar D_B \Xi^B_A = 0.
\end{equation}
\begin{theorem}
Any solution of Einstein's equations with negative cosmological
constant satisfying to the usual Brown-Henneaux boundary conditions  with $\bar
\gamma_{AB}dx^Adx^B = -d\tau^2 + d\phi^2$
will also satisfy to the generalized boundary conditions
\eqref{eq:condasymp1}-\eqref{eq:condasymp4}. 
\end{theorem}
In that sense, we can say that our new boundary conditions are a
generalization of the usual one: we are not losing any solutions.
Using light-cone coordinates, we can put the EOM \eqref{eq:EOMfinal}
in the simple form
\begin{equation}
\label{eq:EOMads3lc}
\d_+\d_- \varphi = 0, \quad \d_+ \Xi^+_- = 0, \quad \d_- \Xi^-_+=0.
\end{equation}
Those quantities for the BTZ black-hole are given by:
\begin{equation}
\label{eq:valuesBTZ}
\varphi=0, \quad \Xi^-_+= 2Gl^2 \left(M + \frac{J}{l} \right), \quad \Xi^+_-= 2Gl^2 \left(M - \frac{J}{l} \right).
\end{equation}

\vspace{3mm}

We would like to emphasize the difference between the two
approaches. In the Brown-Henneaux boundary conditions, one imposes
$\varphi =0$. The EOM then imply $\rho=0$ and $D_B \Xi^B_A = 0$. In the
new boundary conditions, one imposes $\rho = 0$ which leads the EOM
\eqref{eq:EOMfinal} which are EOM for both $\varphi$ and $\Xi^B_A$. We
have new degrees of freedom in $\varphi$.

\section{Surface charges}
\label{sec:surface-charges}

For the surface charges, we follow \cite{Barnich:2002fk}, up to a global change of
sign. The technique allows us to compute the variation of the surface
charges $\ndelta\cQ_\xi[h,g]$ associated to a vector field $\xi$ under a
variation of the metric $h_{\mu\nu} = \delta g_{\mu\nu}$. When this
variation is integrable on field space \cite{Barnich:2008uq}, we can define the charges as:
\begin{equation}
\cQ_\xi[g] = \int_{\gamma_s}\ndelta\cQ_\xi[\delta^s g, g(\gamma_s)]
\end{equation}
where the integration is done along a path $\gamma_s$ in field space
joining the background metric $\bar g$ to the metric $g$ that we
are considering.
For the variation of the charge, we use the expression
\begin{multline}
  \label{eq:291}
  \ndelta\cQ_\xi[\delta g, g]= \int_{\d\Sigma}\frac{\sqrt{- g}}{16
    \pi G}\,(d^{n-2}x)_{\mu\nu}\, \Big[\xi^\nu \nabla^\mu h
  -\xi^\nu \nabla_\sigma h^{\mu\sigma} +\xi_\sigma \nabla^\nu
  h^{\mu\sigma}
  \\
  +\frac{1}{2}h \nabla^\nu\xi^\mu +\half
  h^{\nu\sigma}(\nabla^\mu\xi_\sigma-\nabla_\sigma\xi^\mu)
  -(\mu\leftrightarrow \nu)\Big],
\end{multline}
where the indices are raised and lowered with the full metric $g_{\mu\nu}$
and $\nabla_\mu$ is the covariant derivative
associated to it. We also define
\[(d^{n-k}x)_{\mu\nu}=\frac{1}{k!(n-k)!}
\epsilon_{\mu\nu\alpha_1\dots\alpha_{n-2}}
dx^{\alpha_1}\wedge\dots\wedge dx^{\alpha_{n-2}},\quad
\epsilon_{01\dots n-1}=1,\] with $n=3$ and the surface of integration
$\partial\Sigma$ is taken to be a circle on the cylinder at spatial infinity $C_\infty$. 

A straightforward computation leads to
\begin{multline}
\ndelta\cQ_{\xi}[h,g]=\frac{1}{16\pi G l} \delta\int_{C_\infty} \epsilon_{AB}d x^B \,
\sqrt{-\bar \gamma}\Big[l^2 \bar \gamma^{AC} (\d_C\psi \varphi-\psi \d_C  \varphi
)+ 2 Y^A \bar\gamma^{CD} C_{CD} - 2 Y^C
\bar\gamma^{AD} C_{CD}\Big]\\+\frac{1}{16\pi G l} \int_{C_\infty} \epsilon_{AB}d x^B \,
\sqrt{-\bar \gamma}Y^A \left(\frac{1}{l^2} e^{2\varphi}\delta C_{rr} - 2
  \delta \varphi \bar\gamma^{CD}C_{CD} \right) 
\end{multline}
where we used $\epsilon_{AB} = -\epsilon_{rAB}$. The second line
contains a non-integrable term that is easily removed
using the variation of the constraint \eqref{eq:suppcond}. The
final result, after integrating in field space, is given by: 
\begin{multline}
\label{eq:charges}
\cQ_{\xi}[g]=\frac{1}{16\pi G l} \int_{C_\infty} \epsilon_{AD}d x^D \,
\sqrt{-\bar \gamma}\Big[l^2 (\bar D^A\psi \varphi-\psi \bar D^A  \varphi
)- 2 Y^B\Xi^A_B\Big],
\end{multline}
where we raised and lowered indices with $\bar \gamma_{AB}$ and its inverse.
We see that the 2 relevant
dynamical quantities for the charges are, as expected, $\varphi$ and
$\Xi^A_B$. We normalized our charges using the BTZ black-hole with
$M=0=J$ as a background (or, using \eqref{eq:valuesBTZ}, $\varphi=0$
and $\Xi^A_B=0$). The charges depend only on the leading orders 
of $\xi$: the pure gauge transformations $\xi^r=O(r^{-1})$ and
$\xi^A=O(r^{-4})$ give zero.

In the asymptotically $AdS_3$ case, with $\bar \gamma_{AB}dx^Adx^B =
-d\tau^2 + d\phi^2$ and $C_\infty$ being the circle at $\tau$
constant, we obtain:
\begin{multline}
\label{eq:chargesADS}
\cQ_{\xi}[g]=\frac{1}{16\pi G l} \int_0^{2\pi} d\phi \,
\Big[l^2 (\psi \d_\tau  \varphi-\d_\tau\psi \varphi)+ 2Y^\tau
\Xi_{\tau\tau} + 2 Y^\phi \Xi_{\tau\phi}\Big].
\end{multline}
The last two terms are the usual contribution coming from the two
Virasoro algebras.
Using the time translation and angular rotation symmetry vectors $Y =
\frac{1}{l}\d_\tau$ and $Y = \d_\phi$, we can evaluate the mass and
the angular momentum of the BTZ black-hole:
\begin{equation}
\cQ_{\frac{1}{l}\d_\tau}[g_{BTZ}] = M, \quad \text{and} \quad \cQ_{\d_\phi}[g_{BTZ}] = J.
\end{equation}
In light-cone coordinates and using the parametrization of the
asymptotic symmetry group in term of $(Y^\pm, W^\pm, V)$ introduced in
section \ref{sec:asympt-sym}, the charges \eqref{eq:chargesADS} can be
rewritten as
\begin{multline}
\label{eq:chargesadslc}
\cQ_{\xi}[g]\approx\frac{1}{8\pi G l} \int_0^{2\pi} d\phi \,
\Big[l^2 \left(W^+\d_+ \varphi^+ + W^-\d_- \varphi^- \right)  + Y^+
\Xi_{++} + Y^- \Xi_{--}\Big]\\+ \frac{\alpha}{16\pi G l}\int_0^{2\pi} d\phi \, (W^++W^-)
- \frac{V}{16\pi G l}\int_0^{2\pi} d\phi \,  (\varphi^+ + \varphi^-).
\end{multline}
To obtain this result, we used some integrations by parts and  solved
the EOM \eqref{eq:EOMads3lc} with $\varphi =
\varphi^+(x^+) + \varphi^-(x^-) + \alpha\tau$, $\alpha$ being a constant.

\section{Centrally extended algebra}
\label{sec:algebra}

As in \cite{Barnich:2002fk,Barnich:2008uq}, we expect the charges built in the previous section to
form a representation of the asymptotic symmetry algebra, or more precisely,
that 
\begin{gather}
 \Big[\cQ_{\xi_1}[g], \cQ_{\xi_2}[g] \Big] \equiv \ndelta \cQ_{\xi_1}[\cL_{\xi_2}
  g, g] \approx \cQ_{[\xi_1,\xi_2]_M}[ g]+
K_{\xi_1 , \xi_2},\label{bracket1}
\end{gather}
where $K_{\xi_1 , \xi_2}$ is a possible central extension.

For vectors satisfying \eqref{eq:FGvect}, 
$\cL_\xi g_{AB}$ leads to the following variations:
\begin{eqnarray}
\label{eq:varvarphi}
\delta_\xi \varphi & = & Y^A \d_A \varphi + \frac{1}{2} \bar D_A Y^A -
\frac{1}{2} \psi,\\
\label{eq:varXi}
\delta_\xi \Xi_{AB} & = &
\cL_Y \Xi_{AB} -
\frac{l^2}{2} \bar D_A \d_B \psi \nonumber\\ && + \frac{l^2}{2} \left(
  \d_A \varphi \d_B \psi + \d_B \varphi \d_A \psi - \bar \gamma_{AB} \bar
  D^C \varphi \d_C \psi\right).
\end{eqnarray}
Using those in \eqref{eq:charges} and integration by parts, we get
\begin{multline}
\ndelta \cQ_{\xi_1}[\cL_{\xi_2} g, g] =  \frac{1}{16\pi G l} \int_{C_\infty} \epsilon_{AD}d x^D \,
\sqrt{-\bar \gamma}\Big[l^2 (\bar D^A\hat\psi \varphi-\hat\psi \bar D^A  \varphi
)- 2 \hat Y^B \Xi^A_B\\
+\frac{l^2}{2}\left(\psi_1 \bar D^A \psi_2 - \psi_2 \bar D^A \psi_1
\right) + l^2 \left( Y^B_1 \bar D^A \d_B \psi_2 - Y^B_2 \bar D^A \d_B \psi_1\right)\\
-l^2 \psi_1 Y^A_2 \bar D_B \bar D^B \varphi + \frac{l^2}{2} Y^A_2
\psi_1 \bar R - 2 Y^B_1 Y^A_2 \bar D_E \Xi^E_B\Big].
\end{multline}
On shell, this reproduces \eqref{bracket1} with
\begin{eqnarray}
\label{eq:centralextens}
K_{\xi_1 , \xi_2} &=&  \frac{l}{16\pi G } \int_{C_\infty} \epsilon_{AD}d x^D \,
\sqrt{-\bar \gamma}\Big[\frac{1}{2}\psi_1 \bar D^A \psi_2 + Y^B_1 \bar
D^A \d_B \psi_2 - (1 \leftrightarrow 2)\Big],
\end{eqnarray}
which satisfies the cyclic identity:
\begin{equation}
K_{[\xi_1,\xi_2]_M,\xi_3}+K_{[\xi_2,\xi_3]_M,\xi_1}+K_{[\xi_3,\xi_1]_M,\xi_2}
= 0.
\end{equation}
As expected, the algebra closes and we obtain a non-zero central
extension. However, as one can see clearly in the expression
\eqref{eq:centralextens}, there are no central terms in the conformal
subalgebra parametrized by $Y^A$.

\vspace{5mm}

In the asymptotically $AdS_3$ case, using equation
\eqref{eq:chargesadslc} for the charges and some integration by parts, we obtain
\begin{equation}
\begin{gathered}
  \Big[\cQ_{\xi_1}[g], \cQ_{\xi_2}[g] \Big] \approx
  \cQ_{[\xi_1,\xi_2]_M}[ g]
+ K^+_{\xi_1 , \xi_2} + K^-_{\xi_1 , \xi_2} + K^0_{\xi_1 , \xi_2} ,\\
  K^+_{\xi_1 , \xi_2} = \frac{l}{16\pi G } \int_0^{2\pi}d \phi \,
  (W_1^+\d_+^2Y_2^+ - W_2^+ \d_+^2 Y_1^+ - W_1^+ \d_+ W^+_2),\\
  K^-_{\xi_1 , \xi_2} = \frac{l}{16\pi G } \int_0^{2\pi}d \phi \,
  (W_1^-\d_-^2Y_2^- - W_2^- \d_-^2 Y_1^- - W_1^- \d_- W^-_2),\\
  K^0_{\xi_1 , \xi_2} = \frac{l V_1}{32\pi G } \int_0^{2\pi}d \phi \,
  (W^+_2 + W^-_2) - \frac{l V_2}{32\pi G } \int_0^{2\pi}d \phi \,
  (W^+_1 + W^-_1),
\end{gathered}
\end{equation}
where $K^{\pm,0}_{\xi_1 , \xi_2}$ are the central extensions. Expanding
this result in term of the charges $(L^\pm_m, P^\pm_m, Q)$ associated to
the basis $(l^\pm_m, p^\pm_m, q)$ introduced in section \ref{sec:asympt-sym}, we obtain
explicitly
\begin{equation}
\label{eq:adsfullalgebra}
\boxed{\begin{array}{rclcrcl}
i\left[L^\pm_m,L^\pm_n\right]&=&(m-n)L^\pm_{m+n}, & \qquad & i[L^+_m,L^-_n]&=&0, \\
i\left[L^\pm_m,P^\pm_n\right]&=&-n P^\pm_{m+n} + \frac{l}{8G} im^2\delta_{m+n,0}, & \qquad &  i[L^\pm_m,P^\mp_n]&=&0, \\
i\left[P^\pm_m,P^\pm_n\right]&=&-\frac{l}{8G} m\delta_{m+n,0}, &
\qquad &  i[P^+_m,P^-_n]&=&0, \\
i\left[L^\pm_m, Q\right] & = & \frac{i}{2}P^\pm_m, &\qquad &
i\left[P^\pm_m, Q \right] & = & -i\frac{l}{16 G}\delta_{m,0}.
\end{array}}
\end{equation}
As we saw earlier, adding dynamics to the conformal factor of the boundary metric
sends the Virasoro central charges to zero. This effect is similar to the one described in
\cite{Carlip:2001kk} where Liouville theory is coupled to gravity in two dimensions.

However, as we will see in the next section, there are more than one
2D conformal algebra hidden in this algebra
and it is possible to recover the usual Brown-Henneaux central extension.

\section{Brown-Henneaux central charges recovered}
\label{sec:brown-henn-centr}

At first sight, the final algebra \eqref{eq:adsfullalgebra} is not
very promising. The central charges in the Virasoro algebras are a
key point of the various results obtained in asymptotically $AdS_3$
space-times and we lost them. However, as the boundary conditions
studied are a generalization of the usual ones, the central charges must
be hidden somewhere.

The answer comes by studying the exact Killing vectors of the
background $AdS_3$. The original Virasoro algebras studied by
Brown-Henneaux are built on the Killing vectors of $AdS_3$, in the
sense that, $l^\pm_{-1}, l^\pm_0$ and $l^\pm_{1}$ are the generators
of the $so(2,2)$ algebra leaving $AdS_3$ invariant. As we will show in
the following, the Virasoro generators present in the basis 
used to write our algebra \eqref{eq:ads3alg} do not satisfy this
property. Nevertheless, it is possible to recover it by doing a change of basis of the
algebra. This will also reproduce the usual Brown-Henneaux result for
the central extension of the 2D conformal subalgebra.

The $AdS_3$ metric is given by:
\begin{equation}
ds^2 = -(\frac{r^2}{l^2} + 1) dt^2 + \frac{1}{\frac{r^2}{l^2} + 1}
dr^2 + r^2 d\phi^2.
\end{equation}
In our asymptotic expansion, it corresponds to:
\begin{equation}
\varphi = 0, \quad \Xi_{++} =-\frac{l^2}{4}, \quad \Xi_{--} =-\frac{l^2}{4}.
\end{equation}
The Killing vectors of $AdS_3$ are asymptotic symmetries that preserve those three
quantities. Using the variations \eqref{eq:varvarphi} and
\eqref{eq:varXi}, we obtain the following equations:
\begin{eqnarray}
\delta_\xi \varphi & = & \d_+ Y^+ + \d_- Y^- - \psi = 0, \\
\delta_\xi \Xi_{++} & = & -\frac{l^2}{2}\left(\d_+Y^+ + \d_+^2 \psi
\right) = 0, \\
\delta_\xi \Xi_{--} & = & -\frac{l^2}{2}\left(\d_-Y^- + \d_-^2 \psi
\right) = 0.
\end{eqnarray}
It is obvious that the $l^\pm_{\pm 1}$ defined
in section \ref{sec:asympt-sym} are not solutions to those equations:
they are not Killing vectors of $AdS_3$. The general solution is given
by the set of vectors $(Y^A, \psi=\d_A Y^A)$ with
$Y^A$ satisfying $\d_\pm Y^\pm = \d_\pm^3 Y^\pm$. In terms of our
generators $(l^\pm_m, p^\pm_m)$ the Killing vectors of $AdS_3$ are given by:
\begin{equation}
l^\pm_0, \quad l^\pm_{1} + ip^\pm_1, \quad l^\pm_{-1}-ip^\pm_{-1}.
\end{equation}

We can build two full Virasoro algebras on
those vectors as follows:
\begin{equation}
\tilde l^\pm_m \equiv l^\pm_m + im p^\pm_m.
\end{equation}
The generators $(\tilde l^\pm_m, p^\pm_m, q)$ form a new basis of our
asymptotic symmetry algebra for which the commutators \eqref{eq:ads3alg} take the same form:
\begin{equation}
\begin{array}{rclcrcl}
i[\tilde l^\pm_m,\tilde l^\pm_n]&=&(m-n)\tilde l^\pm_{m+n}, & \qquad & i[\tilde l^+_m,\tilde l^-_n]&=&0, \\
i[\tilde l^\pm_m,p^\pm_n]&=&-n p^\pm_{m+n}, & \qquad &  i[\tilde l^\pm_m,p^\mp_n]&=&0, \\
i[p^\pm_m,p^\pm_n]&=&0, & \qquad & i[p^+_m,p^-_n]&=&0, \\
i[\tilde l^\pm_m,q]&=&\frac{i}{2} p^\pm_m, & \qquad & i[p^\pm_m,q]&=&0.
\end{array}
\end{equation}
On the level of the associated charges $(\tilde L^\pm_m = L^\pm_m
+ im P^\pm_m, P^\pm_m, Q)$, we recover the usual result for the Virasoro
central charges:
\begin{equation}
\label{eq:finalads3algebraever}
\boxed{\begin{array}{rclcrcl}
i[\tilde L^\pm_m,\tilde L^\pm_n]&=&(m-n)\tilde L^\pm_{m+n} +
\frac{c^\pm}{12} m^3 \delta_{m+n,0}, & \qquad & i[\tilde L^+_m,\tilde L^-_n]&=&0, \\
i[\tilde L^\pm_m,P^\pm_n]&=&-n P^\pm_{m+n}, & \qquad &  i[\tilde L^\pm_m,P^\mp_n]&=&0, \\
i[P^\pm_m,P^\pm_n]&=&k \, m\delta_{m+n,0}, & \qquad &  i[P^+_m,P^-_n]&=&0, \\
i[\tilde L^\pm_m, Q] & = & \frac{i}{2}P^\pm_m, &\qquad &
i\left[P^\pm_m, Q \right] & = & i\frac{k}{2}\delta_{m,0}.
\end{array}}
\end{equation}
This central extension is a particular case of the general central
extension studied in appendix \ref{sec:appendix}. Here, only 3 of the 6
possible central charges are non-zero: the
Brown-Henneaux central charges $c^\pm=\frac{3l}{2G}$ and one new quantity
$k=-\frac{l}{8G}$.  The factor of $m^3$ is coming from
our normalization for $\tilde L^\pm_0$: using $AdS_3$ as a background
would lead to the standard $m^3-m$.

Similar algebras appear in the study of higher spin gravity in three
dimensions \cite{Castro:2012bc,Afshar:2012hc}. As in their case and in
the result of \cite{Compere:2013bya}, the central charges $c^\pm$ and $k$ have opposite
signs which, in general, leads to non unitary representations \cite{Detournay:2012pc}.

\section{Conclusions}
\label{sec:conclusions}

The boundary conditions studied in this paper are a generalization of
the usual Brown-Henneaux boundary conditions. Those extended boundary
conditions lead to a second enhancement of the asymptotic symmetry
algebra.  Up to the zero modes, it is generated by two Virasoro algebras and two $U(1)$ current algebras. At the level of the
charges, the algebra is centrally extended and we recover the
Brown-Henneaux central charges $c^\pm=\frac{3l}{2G}$ plus one new
number $k=-\frac{l}{8G}$. In general, the negative value for $k$ leads
to non unitary representations. 

Those boundary conditions give us two interesting things. There are
more degrees of freedom which would maybe account for what we are
missing in our understanding of gravity in three dimensions. The
second improvement is a bigger symmetry algebra. This would give us
more tools to control and understand the theory.

For the future, it would be interesting to see how the results
obtained in the study of asymptotically $AdS_3$ space-times
change. As most of those results rely heavily on
the result of Brown-Henneaux, a change in boundary conditions can have
a strong impact. 

The new chiral boundary conditions of \cite{Compere:2013bya} describe
a different problem than those presented here. One way of seeing it is
that the time translation symmetry is the zero mode of a $U(1)$
current algebra in their case whereas it is part of the
conformal algebra in this generalized Brown-Henneaux case. However, there are still striking
differences in the
number of degrees of freedom and in the size of the algebra. It might
be possible to generalize the chiral boundary 
conditions to allow for more degrees of freedom and maybe enhance the
associated asymptotic symmetry algebra.

\section*{Acknowledgements}
\label{sec:acknowledgements}


I would like to thank G.~Barnich, S.~Detournay, H.~Gonz\'alez, A.~Perez, P.~Ritter, D.~Tempo,
R.~Troncoso and J.~Zanelli for useful discussions. The Centro de Estudios Cient\'ificos
(CECs) is funded by the Chilean Government through the Centers of
Excellence Base Financing Program of Conicyt.

\appendix

\section{Central Extension}
\label{sec:appendix}

Let's consider an algebra $\cG$ generated by $T_a$:
\begin{equation}
\left[T_a, T_b\right] = f^c_{ab} T_c.
\end{equation}
A central extension of $\cG$ by an abelian algebra of dimension 1 is given by a set
of complex numbers $K_{a,b}=-K_{b,a}$ such that the following extended
algebra closes: 
\begin{eqnarray}
\left[T_a, T_b\right] &=& f^c_{ab} T_c + K_{a,b}I,\\
\left[T_a, I\right] &=& 0,
\end{eqnarray}
where $I$ is the new abelian generator \cite{Brink:1988nx,Az:1995nx}.
As it is customary, we will not write it in
the rest of the computation. Two such central extensions are
equivalent if they can be related by a redefinition of the
generators of $\cG$ : $T_c \rightarrow T_c + \alpha_c I$. 

\vspace{3mm}

To compute the most general central extension of the algebra
(\ref{eq:ads3alg}) up to equivalence, we will start with the general form:
\begin{equation}
\label{eq:extealgebraever}
\begin{array}{rclcrcl}
i[L^\pm_m, L^\pm_n]&=&(m-n) L^\pm_{m+n} +
K^\pm_{m,n}, & \qquad & i[ L^+_m,L^-_n]&=&K^{+-}_{m,n}, \\
i[ L^\pm_m,P^\pm_n]&=&-n P^\pm_{m+n} + V^\pm_{m,n}, & \qquad &  i[L^\pm_m,P^\mp_n]&=&V^{\pm\mp}_{m,n}, \\
i[P^\pm_m,P^\pm_n]&=&W^\pm_{m,n}, & \qquad &  i[P^+_m,P^-_n]&=&W^{+-}_{m,n}, \\
i[ L^\pm_m, Q] & = & \frac{i}{2}P^\pm_m + X^\pm_m, &\qquad &
i\left[P^\pm_m, Q \right] & = & Y^\pm_m,
\end{array}
\end{equation}
with $K^\pm_{m,n}=-K^\pm_{n,m}$ and $
W^\pm_{m,n}=-W^\pm_{n,m}$. Using redefinitions of the
generators:
\begin{eqnarray}
L^\pm_n & \rightarrow & L^\pm_n - \frac{1}{n} K^\pm_{0,n} \qquad
\text{for}\quad n\ne0,\\
L^\pm_0 & \rightarrow & L^\pm_0 + \frac{1}{2} K^\pm_{1,-1},\\
P^\pm_n & \rightarrow & P^\pm_n - \frac{1}{n} V^\pm_{0,n} \qquad
\text{for}\quad n\ne0,\\
P_0 & \rightarrow & P_0 - i (X^+_0 + X^-_0),
\end{eqnarray}
we can put the following quantities to zero: $K^\pm_{0,m}$,
$K^\pm_{1,-1}$, $V^\pm_{0,n}$ for $n\ne0$ and $X^+_0 + X^-_0$. Because
the generator $Q$ never appears on the right hand side, a redefinition
of $Q$ will not produce any central term: we have used all our freedom.

\vspace{3mm}

The extended algebra (\ref{eq:extealgebraever}) is an algebra if and
only if it
satisfies the Jacobi identity. Let's check this step by step:
\begin{itemize}
\item The various Jacobi identities that we can
  write with $L^\pm_m$ give the following equations:
\begin{eqnarray}
0 & = & (m-n) K^\pm_{m+n,p} + (n-p) K^\pm_{n+p,m}  + (p-m)
K^\pm_{p+m,n},\\
0 & = & (m-n) K^{+-}_{m+n,p}.
\end{eqnarray}
Using the fact that $K^\pm_{0,m} = K^\pm_{1,-1} = 0$, one can easily
prove that:
\begin{equation}
K^\pm_{m,n} = \frac{c^\pm}{12}(m^3-m)\delta_{m+n,0}, \qquad
K^{+-}_{m,n}=0,
\end{equation}
which is the usual result. The two numbers $c^\pm$ are arbitrary and
are the two central charges of the conformal group in 2D. Using
another choice for the redefinition 
of $L^\pm_0$, $L^\pm_0 \rightarrow L^\pm_0 - \frac{c^\pm}{24}$, we
can put $K^\pm_{m,n} = \frac{c^\pm}{12}m^3\delta_{m+n,0}$.
\item The Jacobi identities involving two Virasoro generators
  $L^\pm_m$ and one current generator $(P^\pm_m, Q)$ give:
\begin{eqnarray}
0 & = & (m-n)V^\pm_{m+n,p} + p(V^\pm_{m,n+p}-V^\pm_{n,m+p}),\\
0 & = & (m-n) V^{\pm\mp}_{m+n,p},\\
0 & = & (m-n) X^\pm_{m+n} - \frac{i}{2} V^\pm_{m,n} + \frac{i}{2} V^\pm_{n,m}.
\end{eqnarray}
This time, the solution is parametrized by 3 numbers $d^\pm$
and $d_0$:
\begin{equation}
V^\pm_{m,n} = \left(d^\pm m^2 \mp i d_0 \,m(m+2)\right)\delta_{m+n,0},
\quad X^\pm_m = \pm d_0 \, \delta_{m,0}, \quad V^{\pm\mp}_{m,n}=0.
\end{equation}
\item The Jacobi identities involving only one Virasoro generator
  and two current generators give:
\begin{eqnarray}
0 & = & -n W^\pm_{m+n,p} + p W^\pm_{m+p,n},\\
0 & = & -n W^{\pm\mp}_{m+n,p},\\
0 & = & -n Y^\pm_{m+n} - \frac{i}{2} W^\pm_{m,n}.
\end{eqnarray}
Because $P^+_0$ and $P^-_0$ represent the same generator, we have
$Y^+_0=Y^-_0$. This restricts the solution to
\begin{equation}
W^\pm_{m,n} = k m\, \delta_{m+n,0}, \qquad W^{\pm\mp}_{m,n}=0, \qquad Y^\pm_m =\frac{i}{2}k\, \delta_{m,0}, 
\end{equation}
which is parametrized by only one number: $k$.
\item The Jacobi identities involving only current generators are
  automatically satisfied.
\end{itemize}

\vspace{3mm}

The final result is then that, up to redefinition of the
generators, the most general central extension of
the algebra (\ref{eq:ads3alg}) is parametrized by $6$ numbers $c^\pm$,
$d^\pm$, $d^0$ and $k$: 
\begin{equation}
\label{eq:appextealgebra}
\begin{array}{rclcrcl}
i[L^\pm_m, L^\pm_n]&=&(m-n) L^\pm_{m+n} +
\frac{c^\pm}{12}m^3 \delta_{m+n,0}, & \qquad & i[ L^+_m,L^-_n]&=& 0, \\
i[ L^\pm_m,P^\pm_n]&=&-n P^\pm_{m+n} + \left(d^\pm m^2 \mp id_0 \, m(m+2) \right) \delta_{m+n,0}, & \qquad &  i[L^\pm_m,P^\mp_n]&=&0, \\
i[P^\pm_m,P^\pm_n]&=& k m \delta_{m+n,0} , & \qquad &  i[P^+_m,P^-_n]&=& 0, \\
i[ L^\pm_m, Q] & = & \frac{i}{2}P^\pm_m \pm d_0 \, \delta_{m,0}, &\qquad &
i\left[P^\pm_m, Q \right] & = & \frac{i}{2} k \delta_{m,0}.
\end{array}
\end{equation}

\bibliography{./biblio.bib}

\end{document}